\documentclass{ws-procs9x6}
\usepackage{graphicx}
\usepackage{epstopdf}
\date{today}                                         
\begin{document}
\title{STRANGENESS PHYSICS WITH CLAS AT JLAB\\}
\author{Volker D. Burkert}
\address{Jefferson Lab, Newport News, VA 23606, USA\\$^*$E-mail: burkert@jlab.org\\}                    
\begin{abstract}
A brief overview of strangeness physics with the CLAS detector at JLab is given, mainly covering the 
domain of nucleon resonances. Several excited states predicted by the symmetric 
constituent quark model may have significant couplings to the 
$K\Lambda$ or $K\Sigma$ channels. I will discuss data that are relevant in the search for such states in 
the strangeness channel, and give an outlook on the future prospects of the N* program at JLab with 
electromagnetic probes.
\end{abstract}
\section{Introduction}\label{aba:intro}
A major goal of hadron physics with electromagnetic probes is to study the 
structure of the nucleon and its excited states. 
The nucleon excitation spectrum is a direct reflection of the underlying 
degrees of freedom. For example, the $SU(6)$ symmetric 
constituent quark model predicts a large number of states many of which have 
not been observed experimentally using hadronic probes. A model that 
includes clustering of two quarks
into di-quarks has fewer degrees of freedom and predicts a smaller 
number of excited states. It is obviously important to obtain a better understanding of 
the degrees of
freedom underlying the nucleon properties. Most studies of the nucleon spectrum 
have used pion probes. Electromagnetic probes, photons and electrons, and 
strangeness production are complementary ways to search for some of the excited
states and can help discriminate between alternative description of the nucleon 
spectrum.  The CLAS program is designed to accurately measure electromagnetic 
cross sections and single and double polarization observables with wide energy 
and angle coverage. 
\section{Experimental aspects}
The experimental program makes use of the CEBAF Large Acceptance Spectrometer (CLAS)
~\cite{clas} which provides particle identification and
momentum analysis in a polar angle range from 8$^\circ$ to 140$^\circ$. The 
photon energy tagger provides energy-marked photons with an energy resolution of 
${\sigma(E) \over E} = 10^{-3}$. Other equipment includes a coherent 
bremsstrahlung facility with a goniometer for diamond crystal positioning and 
angle control. The facility has been used for linearly polarized photons with 
polarizations up to 90\%. There are two frozen spin polarized 
targets, one based on butanol as target material (FROST), 
and one using HD as a target material (HD-Ice). The latter is currently under construction.
FROST has already been operated successfully in 
longitudinal polarization mode and will be used in transverse polarization 
mode in 2010.
\begin{figure}
\begin{center}
\includegraphics[width=10cm]{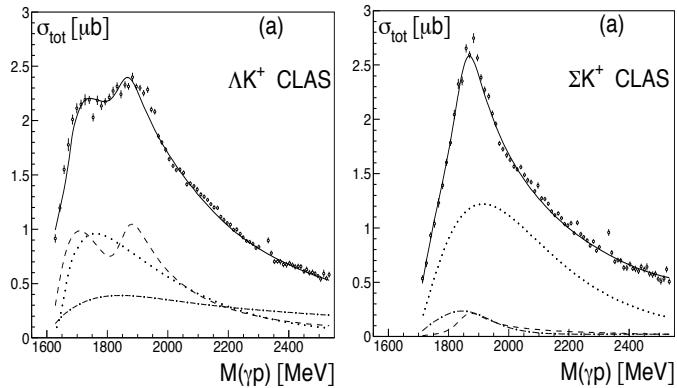}
\end{center}
\caption{Integrated total cross section data for $K^+\Lambda$ and $K^+\Sigma^\circ$ channels. 
The lines represent the fit results of the Bonn-Gatchina coupled-channel analysis. The dashed line shows
the energy-dependence of the $P_{13}$ partial wave which indicates the presence of two $P_{13}$ resonances 
at 1720 MeV and at 1900 MeV, respectively.  The new $P_{13}(1900)$ contributes a significant 
fraction to the total $K^+\Lambda$ cross section.}
\label{fig:lambda_int_cs}
\end{figure}  
HD-Ice will be used as polarized neutron (deuteron) target in 2010/2011. 
Circularly polarized photons can be generated by scattering the highly polarized electron beam from 
an amorphous radiator. It is well understood that measurements of differential cross sections in 
photoproduction of single pseudoscalar mesons alone results in ambiguous solutions 
for the contributing resonant partial waves. The $N^*$ program at JLab is therefore 
aimed at complete, or nearly complete measurements for processes 
$\vec{\gamma} \vec{p} \rightarrow \pi N, ~\eta p,~K^+\vec{Y}$ and 
$\vec{\gamma} \vec{n} \rightarrow \pi N, ~K\vec{Y}$. Complete information can be 
obtained by using a combination of linearly and circularly polarized photon beams, 
measurement of hyperon recoil polarization, and the use targets with longitudinal 
and transverse polarization. The reaction is fully described by 4 complex
 parity conserving amplitudes, requiring 8 combinations of 
beam, target, and recoil polarization measurements for an 
unambiguous extraction of the scattering amplitude. If all possible combinations are 
measured, 16 observables can be extracted. In measurements that involve nucleons 
in the final state where the recoil polarization is not measured, 7 independent observables 
can be obtained directly, and the recoil polarization asymmetry $P$ can be inferred 
from the double polarization asymmetry with linearly polarized beam and transverse target 
polarization. 
\begin{figure}[t]
\begin{center}
\includegraphics[width=10cm]{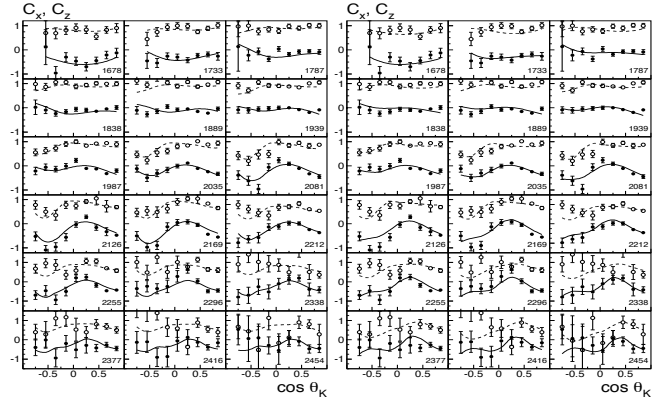}
\end{center}
\caption{Angular dependence of polarization transfer observables $C_{x'}, C_z$. The panel on the left 
and right show different solutions of the Bonn-Gatchina fit. A $P_{13}(1900)$ resonance 
is needed to obtain a good fit. }
\label{fig:lambda_cs}
\end{figure}
\section{Search for $N^*$ states in $KY$ channels} 
A large amount of cross section data have been collected in recent years on the $K\Lambda$ and 
$K\Sigma$ photo-production~\cite{mcnabb04,brad06}. These data cover the nucleon resonance region
in fine steps of about $10$~MeV in the hadronic mass W, and nearly the entire polar angle range. 
The integrated cross section shown in 
Fig.~\ref{fig:lambda_int_cs} reveals a strong bump around $W=1900$~MeV which may be the result of s-channel
resonance production. However, more definite conclusions can only be drawn when polarization observables
are included in the analysis.  First differential cross sections of the channel $\gamma p \rightarrow K^{*\circ}\Sigma^+$, 
which probes the mass range $W > 2100$~MeV have been measured recently~\cite{hleiqawi07}, and will be continued 
at higher statistics. 
\begin{figure}
\begin{center}
\includegraphics[width=9cm]{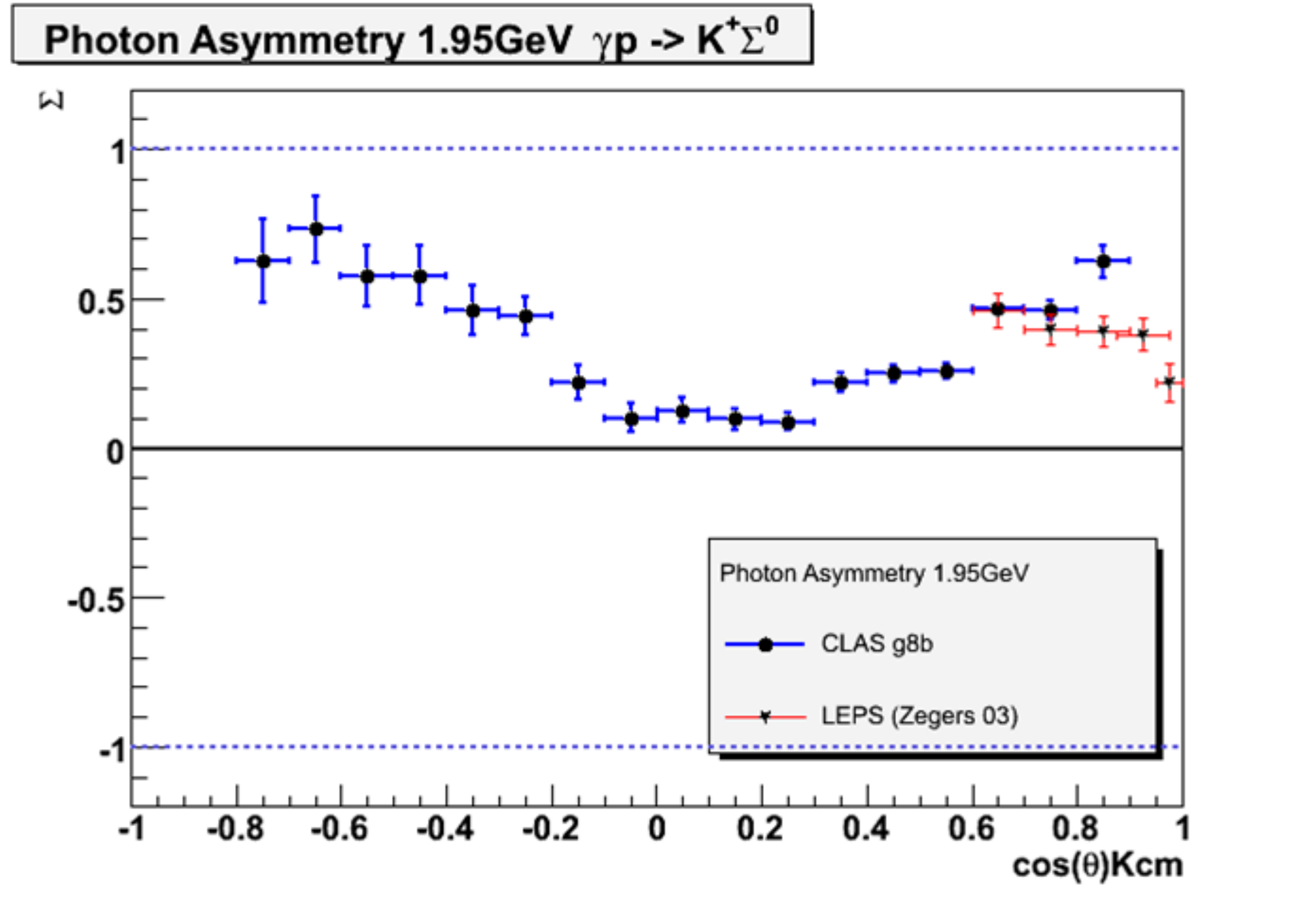}
\end{center}
\caption{Preliminary data from CLAS for the beam asymmetry of the $K\Sigma^\circ$ final state.}
\label{fig:g8b_asymmetry}
\end{figure}  
\subsection{Polarized beam and spin transfer} 
\begin{figure}[t]
\begin{center}
\vspace{-1cm}
\includegraphics[width=10cm]{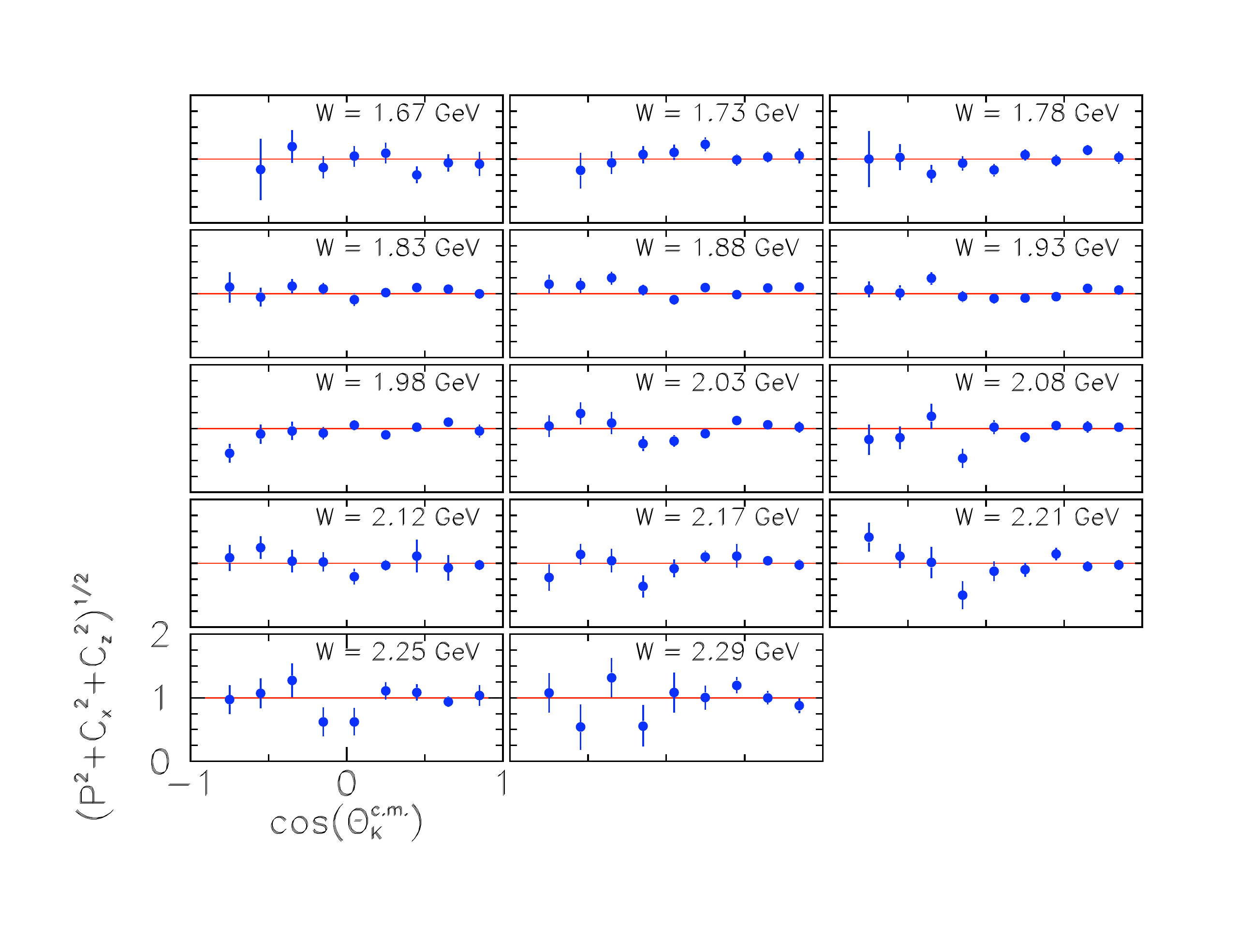}
\end{center}
\caption{Spin transfer from the polarized photon to the final state $\Lambda$. The spin transfer  
data were fitted by the Bonn-Gatchina group simultaneously with the CLAS $K^+\Lambda$ and $K^+\Sigma$ differential 
cross section data. A $P_{13}(1900)$ state is required for the best fit to the cross section 
and spin transfer data.}
\label{fig:lambda_spin}
\end{figure}

Fitting differential cross sections alone has not resulted in 
unambiguous identification of a specific s-channel resonance. In addition to precise $K\Lambda$ and $K\Sigma$
cross section data, recoil polarization and polarization transfer data have been 
measured~\cite{brad07}. The recoil polarization data in the $K^+\Lambda$ sector showed a 
highly unexpected behavior: The spin transfer from the circularly polarized photon to the $\Lambda$ 
hyperon is complete, leaving the $\Lambda$ hyperon 100\% polarized, as can be seen in Fig.~\ref{fig:lambda_spin}. 
The sum of all polarization components $R^2 \equiv P^2 + C_x^2 + C_z^2$ which has an upper bound of 1, 
is consistent with $R^2 = 1$ throughout the region covered by measurement. At first glance this 
result seems adverse to the idea that the $K\Lambda$ final state has a significant component of 
$N^*$ resonance associated with. However, the analysis of the combined CLAS differential 
cross section and polarization 
transfer data by the Bonn-Gatchina group~\cite{nikonov08} shows strong sensitivity to a 
$P_{13}(1900)$ candidate state. The decisive ingredient in this analysis is the CLAS spin transfer 
data set. We remark that 
the peak observed in the $K^+\Lambda$ data seen near 1900 MeV in Fig.~\ref{fig:lambda_cs} 
was originally attributed to a $D_{13}(1900)$
resonance before the spin transfer data became available. A $P_{13}(1900)$ is listed as a 2-star candidate
state in the 2008 edition of the RPP~\cite{pdg2008}. If this assignment is confirmed in future
analyses which should include additional polarization data, the existence of a $P_{13}(1900)$ 
state would be strong evidence against the quark-diquark model. This model has no place for such a state 
in this mass range~\cite{santo05}. 
New precise high statistics data with linearly polarized photon beam on proton targets have also become 
available. Preliminary results for the process $\gamma p \rightarrow K^+\Sigma^\circ$  are shown for a
single energy bin in Fig.~\ref{fig:g8b_asymmetry}. Eventually, these data will span the resonance mass 
region up to $W=2.5$~GeV.  
Several other strangeness channels such as 
$\gamma n \rightarrow K^\circ\Lambda$, $K^{*\circ} \Lambda$, $K^\circ\Sigma^\circ$, $K^+\Sigma^-$, 
and $K^+\Sigma^-(1385)$ are currently being investigated to search 
for new states on neutron targets using circularly and linearly polarized photon beams.  For most of 
these channel a complete kinematical reconstruction in the neutron rest frame is possible, 
thus eliminating the effect of Fermi motion in the deuteron nucleus. Recoil polarization measurements in all 
hyperon final states are also available for the analysis.
\section{Search for new cascade baryons}
\begin{figure}
\begin{center}
\includegraphics[width=9cm]{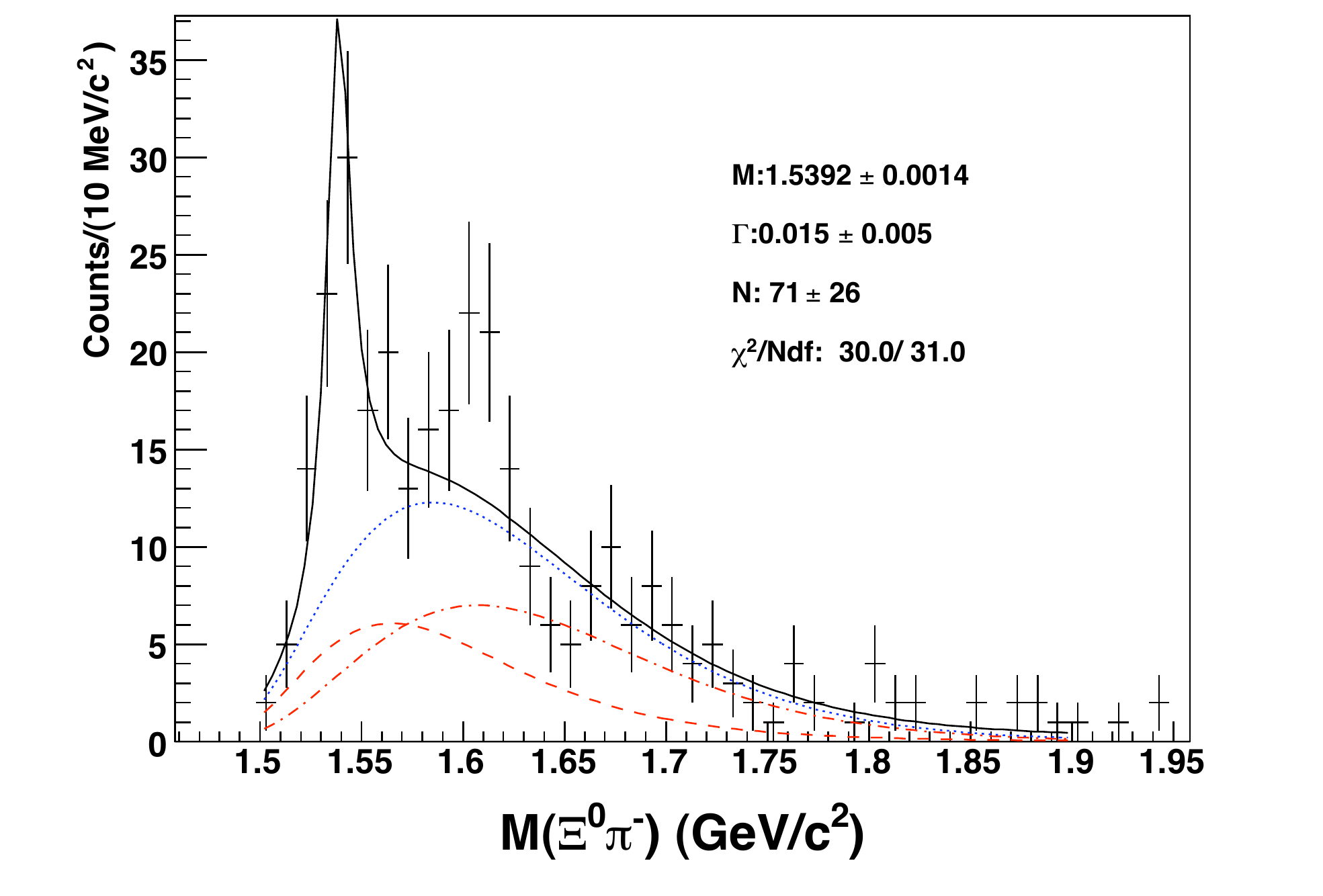}
\end{center}
\caption{Mass of the $\Xi\pi^-$ final state. The $\Xi(1530)$ is clearly seen. 
A structure near 1620 MeV could be related to a dynamically 
generated $\Xi$ state predicted in the model of Ref.~\cite{oset} . }
\label{fig:cascade}
\end{figure} 
The production of $\Xi$ hyperons, i.e. strangeness $S = -2$ excited states, presents another 
promising way of searching for new baryon states. The advantages of cascade hyperon states 
are due to the 
expected narrow widths of theses states compared to $S = 0$, and $S = -1$ resonances. 
The disadvantages of using photon beams are also obvious: The $S=-2$ requires production of 
at least two kaons in the final state. Possible production mechanisms include t-channel $K$ or $K^*$ 
exchanges on proton targets with an excited hyperon $Y^*$ ($\Lambda^*$ or $\Sigma^*$) as 
intermediate state and 
subsequent decays $Y^* \rightarrow K^+ \Xi^*$ and $\Xi^* \rightarrow \Xi \pi$ or $\Xi \rightarrow \Lambda (\Sigma) \bar{K}$.  
Missing mass technique may be used to search for new states in the reaction $\gamma p \rightarrow K^+ K^+ X$ if the state is 
sufficiently narrow to be observed as a peak in the missing mass spectrum. However, an analysis of the 
final state is needed to assign spin and parity to the state. Data from CLAS taken at 3.6 GeV beam energy 
show that  one can 
identify the lowest two cascade states this way.~\cite{guo07}. To identify the higher mass states higher energy is needed. 
\begin{figure}
\begin{center}
\vspace{-1cm}
\includegraphics[width=11cm]{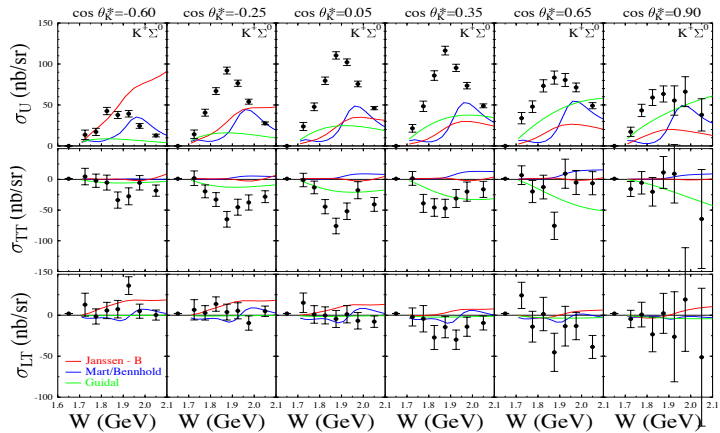}
\end{center}
\vspace{-2cm}
\caption{Separated response functions for $K^+\Sigma^\circ$ electroproduction. The response function $\sigma_u$ 
and $\sigma_{TT}$ show strong resonance-like structure near $W=1.9$GeV, which is not reproduced by any model.}
\label{fig:Sigma0el}
\end{figure}  
Another approach to isolate excited cascades and determine their spin-parity, is to measure 
additional particles in the 
final state. The invariant mass of the $\Xi\pi^-$ system is displayed in Fig.~\ref{fig:cascade} and
shows the first exited state $\Xi(1530)$ and indications of additional structure near 1620 MeV.  A state 
near that mass is predicted~\cite{oset} as a dynamically generated $\Xi\pi$ system. The 
data have insufficient statistics and were taken at too low energy to allow further investigations. New 
data taken at 5.7 GeV electron energy and with higher statistics are currently being analyzed, and should 
allow more definite conclusions on a possible new $\Xi$ state at that mass.
\section{Strangeness electroproduction}
Significant effort has been devoted to the study of electroproduction of hyperons~\cite{ambros07,carman09} as a complementary
 means of searching for new excited nucleon states. Figure~\ref{fig:Sigma0el} shows the dependence of the response functions 
on the hadronic mass W. The comparison with model calculations reveals large discrepancies. None of the models that include
known nucleon resonance couplings to $K\Sigma$, is able to get the normalization correct. This leaves much room for yet to be
identified resonance strength.        
\begin{figure}[h]
\begin{center}
\includegraphics[width=11cm]{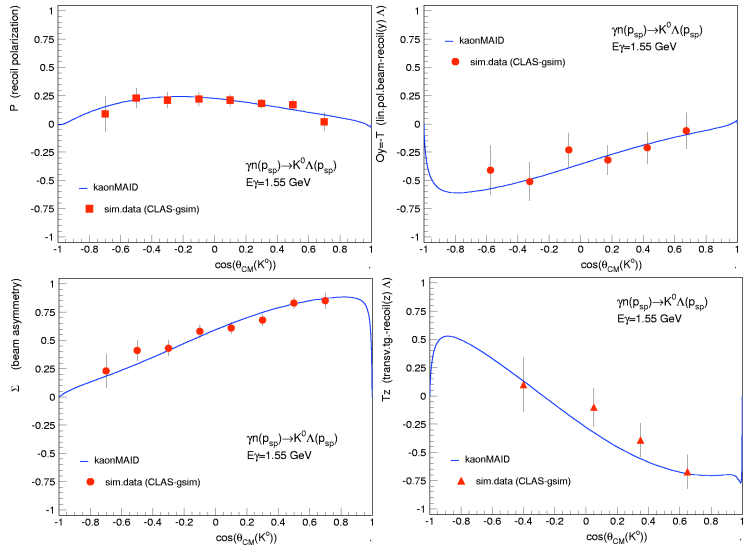}
\end{center}
\caption{Data samples for the process $\gamma \vec{n} \rightarrow K_s^\circ \vec{\Lambda}$ projected 
for the upcoming polarized target run using the HD-Ice facility at JLab. }
\label{fig:HD_proj}
\end{figure}  
\section{Outlook}
The power of polarization measurements for the resonance analysis will be brought to bear when 
double and triple polarization data are available making use of longitudinally and transversely 
polarized proton and neutron targets. Data with the polarized proton target (FROST) 
combined with linearly and circularly polarized photons are currently in the analysis stage. 
Measurements with transverse polarized proton targets are planned for 2010. 

Measurement of polarization observables using polarized neutrons are planned for 2010/2011. Some
projected data using the HD-Ice target facility in the CLAS detector are shown in 
Fig.~\ref{fig:HD_proj}. Single polarization observables $P$, $\Sigma$ and 
double polarization asymmetries for beam-recoil polarization $O_{y'}$, and  target-recoil polarization 
$T_{z'}$ are shown. Other observables, e.g. target asymmetry $T$, beam-target asymmetries $E,~F$, 
and $G,~H$ will be measured as well. The projections are for one photon energy bin out of over 25 bins. 
The solid line is the projection of the kaonMAID code~\cite{kaonmaid}.

\vspace{1cm}

{Authored by The Southeastern Universities Research 
Association, Inc. under U.S. DOE Contract No. DE-AC05-84ER40150 . The U.S. Government 
retains a non-exclusive, paid-up, irrevocable, world-wide license to publish or reproduce 
this manuscript for U.S. Government purposes.}

\end{document}